\documentclass[aps,prl,superscriptaddress,twocolumn]{revtex4}
\usepackage{graphics}
\usepackage{amsmath}
\usepackage{graphicx}
\usepackage{amsfonts}
\usepackage{amssymb}

\begin{document}

\title{Security of two-way quantum cryptography against asymmetric Gaussian
attacks}
\author{Stefano Pirandola}
\affiliation{MIT - Research Laboratory of Electronics, Cambridge MA 02139, USA}
\author{Stefano Mancini}
\affiliation{Dipartimento di Fisica \& CNISM, Universit\`{a} di Camerino, I-62032
Camerino, Italy}
\author{Seth Lloyd}
\affiliation{MIT - Research Laboratory of Electronics, Cambridge MA 02139, USA}
\affiliation{MIT - Department of Mechanical Engineering, Cambridge MA 02139, USA}
\author{Samuel L. Braunstein}
\date{\today }
\affiliation{Computer Science, University of York, York YO10 5DD, United Kingdom}

\begin{abstract}
Recently, we have shown the advantages of two-way quantum communications in
continuous variable quantum cryptography. Thanks to this new approach, two
honest users can achieve a non-trivial security enhancement as long as the
Gaussian interactions of an eavesdropper are independent and identical. In
this work, we consider asymmetric strategies where the Gaussian interactions
can be different and classically correlated. For several attacks of this
kind, we prove that the enhancement of security still holds when the two-way
protocols are used in direct reconciliation.
\end{abstract}

\maketitle

\section{Introduction to continuous variable quantum cryptography}

In recent years, quantum information has discovered the non-trivial
advantages offered by continuous variable systems, i.e., quantum systems
described by a set of observables, like position and momentum, having a
continuous spectrum of eigenvalues \cite{CVbook}. Accordingly, quantum key
distribution\ has been extended to this new framework \cite%
{Ralph1,Preskill,Homo,Hetero} and cryptographic protocols based on coherent
states have been proven to be very powerful for their experimental
feasibility \cite{Homo2,Hetero2}. In these quantum key distribution\
protocols, Alice prepares a coherent state $\left\vert \gamma \right\rangle $
whose amplitude $\gamma =(Q+iP)/2$ encodes two random variables $Q$ and $P$
following two independent Gaussian distributions (having zero mean and the
same large variance). Then, Alice sends the state to Bob, who measures it in
order to retrieve the encoded information. Such a measurement can be:

\begin{description}
\item[(i)] A measurement of $Q$ \textit{or} $P$, randomly chosen by Bob.
Such a disjoint measurement is called homodyne detection and, therefore, we
call \textquotedblleft homodyne\textquotedblright\ ($Hom$) the corresponding
protocol \cite{Homo,Homo2}.

\item[(ii)] A joint measurement of $Q$ \textit{and} $P$. This measurement is
called heterodyne detection and is equivalent to a balanced beam splitter
followed by two homodyne detectors. We call \textquotedblleft
heterodyne\textquotedblright\ ($Het$) the corresponding protocol \cite%
{Hetero,Hetero2}.
\end{description}

\noindent In both protocols, Alice and Bob finally share pairs of correlated
continuous variables. From these variables they can extract a secret binary
key via slicing techniques of the phase space \cite{GaussRec}. This
classical stage is called \emph{reconciliation} and can be \emph{direct} if
Bob estimates Alice's original variables or \emph{reverse} if Alice
estimates Bob's outcomes \cite{Estimators}.

\begin{figure}[tbph]
\vspace{-1.1cm}
\par
\begin{center}
\includegraphics[width=0.52\textwidth] {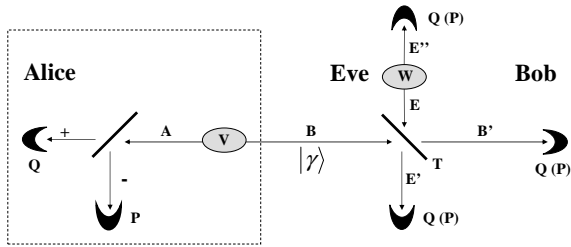}
\end{center}
\par
\vspace{-1.4cm}
\caption{Individual entangling cloner attack against the $Hom$ protocol.
(See text for explanation.) The dashed line displays a black-box, with an
EPR\ source and a heterodyne detector inside, which Alice can use to prepare
a randomly displaced coherent state $\left\vert \protect\gamma \right\rangle
$ .}
\label{Attack1}
\end{figure}

Even if these protocols belong to the so-called prepare and measure (PM)
schemes, they can be equivalently formulated in terms of
Einstein-Podolsky-Rosen (EPR) schemes, where Alice and Bob extract a secret
key from the correlated outcomes of the measurements made upon a shared EPR%
\emph{\ source}. This source is realized by a two-mode squeezed vacuum state
whose correlation matrix is equal to
\begin{equation}
\mathbf{V}=\left(
\begin{array}{cc}
V\mathbf{I} & \sqrt{V^{2}-1}\mathbf{Z} \\
\sqrt{V^{2}-1}\mathbf{Z} & V\mathbf{I}%
\end{array}%
\right) ,  \label{CM_EPR}
\end{equation}%
where $\mathbf{Z}\equiv \mathrm{diag}(1,-1)$, $\mathbf{I}$ the $2\times 2$
identity matrix, and $V$ is a variance characterizing the source \cite%
{QObook}. One can easily show that heterodyning one mode of this EPR\ source
is equivalent to the remote preparation of a coherent state $\left\vert
\gamma \right\rangle $ whose amplitude $\gamma $ is randomly modulated by a
Gaussian of variance $V-1$ (see Appendix). The EPR formulation of the $Hom$
protocol is depicted in Fig.~\ref{Attack1} where the attack of a potential
eavesdropper, Eve, is also shown. According to the standard eavesdropping
scenario, we consider an individual Gaussian attack which is based on the
usage of an entangling cloner \cite{Homo2}. In this attack, each signal sent
from Alice to Bob (mode $B$) is mixed with a \emph{probe} (mode $E$), via a
beam splitter of transmission $T$. This probe is part of an EPR\ source with
variance $W$ which is in Eve's hands. At the end of the protocol, when Bob
reveals the basis ($Q$ or $P$) chosen for each run, Eve will consequently
perform the appropriate homodyne measurements ($Q$ or $P$) on her output
modes $E^{\prime }$ and $E^{\prime \prime }$. From such measurements, Eve
will infer Alice's variable (direct reconciliation) or Bob's variable
(reverse reconciliation). An entangling cloner attack can be therefore
characterized by two parameters, transmission $T$ and variance $W$, which
can be arranged in the unique quantity
\begin{equation}
\Sigma \equiv W(1-T)T^{-1}~,
\end{equation}%
representing the variance of the Gaussian noise added by the channel. These
quantities are evaluated by Alice and Bob by publishing part of their
correlated continuous variables $Q_{B^{\prime }}$ and $Q_{+}$, or $%
P_{B^{\prime }}$ and $P_{-}$ (see Fig.~\ref{Attack1}). In this way, they
perform an error analysis of the channel, which provides the correlation
matrix\ $\mathbf{V}^{\prime }$\ of their shared Gaussian state $\rho
_{AB^{\prime }}$\ and, therefore, their mutual information $%
I_{AB}=I(Q_{B^{\prime }},Q_{+})$ (see Appendix). Similarly, they can
evaluate $I_{AE}$ and $I_{BE}$, and therefore the two key-rates $%
I_{AB}-I_{AE}$ (direct reconciliation) and $I_{AB}-I_{BE}$ (reverse
reconciliation). The security thresholds achieved for high modulation ($%
V\rightarrow +\infty $)\ are equal to $\Sigma =1$ for direct reconciliation,
and to $\Sigma =T^{-1}>1$ for reverse reconciliation. In particular, for
direct reconciliation, the optimal attack is given by an entangling cloner
with $W=1$, i.e., a beam splitter (\emph{lossy channel attack}). In such a
case, the threshold simply corresponds to $T=1/2$, i.e., 3 dB of losses \cite%
{Homo}. Similar results \cite{Hetero} hold for the $Het$ protocol.

\section{Two-way protocols}

\begin{figure}[tbph]
\vspace{-0cm}
\par
\begin{center}
\includegraphics[width=0.35\textwidth] {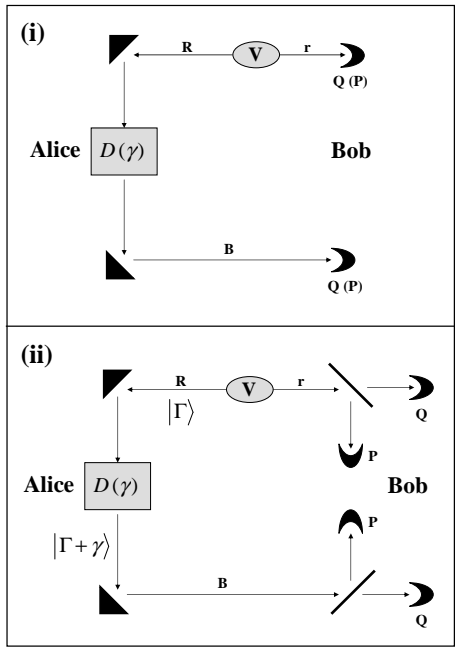}
\end{center}
\par
\vspace{-0.3cm}
\caption{Two-way quantum cryptography. Inset (i) shows the $Hom^{2}$
protocol, i.e., the ON configuration of the hybrid protocol $Hom^{1,2}$.
Inset (ii) shows the $Het^{2}$ protocol, i.e., the ON configuration of the
hybrid protocol $Het^{1,2}$.}
\label{Tutti_MM}
\end{figure}

Even if the underlying physical principles are the same, different protocols
are able to exploit them with different performances. In Ref.~\cite{NatPhys}%
, we have shown that a security enhancement can be achieved by resorting to
a multiple quantum communication (QC) between the trusted parties. In this
approach, a bosonic mode is transmitted forward and backward between the two
parties in order to store and distribute the secret information. Here we
briefly review these protocols and then we study their security by assuming
attacks which are asymmetric between the forward and backward paths. As
depicted in Fig.~\ref{Tutti_MM}, we may consider two different types of
two-way protocols:

\begin{description}
\item[(i)] \textit{Two-way homodyne (}$Hom^{2}$\textit{) protocol.}~The $%
Hom^{2}$ protocol extends the $Hom$ protocol to two-way QC. In the $Hom^{2}$
protocol, Bob has an EPR source (with variance $V$), of which he keeps a
mode $r$ while he sends the other \emph{reference} mode $R$\ to Alice. Then,
Alice randomly displaces this mode in phase-space. This means that she
applies a displacement operator \cite{QObook} $D(\gamma )$ whose amplitude $%
\gamma =(Q+iP)/2$ follows a Gaussian distribution\ with $\left\langle
Q^{2}\right\rangle =\left\langle P^{2}\right\rangle =V$ and $\left\langle
QP\right\rangle =\left\langle Q\right\rangle =\left\langle P\right\rangle =0$%
. The final mode $B$\ is then sent back to Bob. This mode contains Alice's
signal $\gamma $, since its quadratures are equal to $Q_{B}=Q_{R}+Q$ and $%
P_{B}=P_{R}+P$. In order to access this signal, Bob homodynes his modes $r$
and $B$ by choosing to measure their position or momentum at random. For
instance, he can decide to measure positions $Q_{r}$ and $Q_{B}$, so that he
can construct an optimal estimator of $Q_{R}$ (from $Q_{r}$) and, then, an
estimator $Q^{(B)}$ of $Q=Q_{B}-Q_{R}$. Symmetrically, he can measure $P_{r}$
and $P_{B}$ to infer $P$. The basis chosen for each run of the protocol will
be classically communicated to Alice at the end of protocol, when the two
trusted parties will share pairs of correlated continuous variables $%
\{Q,Q^{(B)}\}$ and $\{P,P^{(B)}\}$.

\item[(ii)] \textit{Two-way heterodyne (}$Het^{2}$\textit{) protocol.}~As
for the one-way protocols, Bob can perform a joint measurement of $Q$ and $P$%
. This is achieved in the $Het^{2}$\ protocol which extends the $Het$
protocol to two-way QC. Here, Bob heterodynes his modes $r$ and $B$, from
whose results he infers the full signal $(Q,P)$ of Alice. Notice that this
protocol does not need any final basis revelation. Further, it can be fully
implemented with coherent states. In fact, by heterodyning mode $r$, Bob
equivalently prepares a coherent state $\left\vert \Gamma \right\rangle
=D(\Gamma )\left\vert 0\right\rangle $ which is sent to Alice. This state is
a \emph{reference} state which contains the reference random transformation $%
\Gamma $\ known to Bob. By applying her random displacement $D(\gamma )$,
Alice transforms this state into another coherent state $\left\vert \Gamma
+\gamma \right\rangle $ which is sent back to Bob via the mode $B$. By
subsequent heterodyne detection, Bob is able to estimate the total amplitude
$\Gamma +\gamma $ and, therefore, to infer $\gamma $ from the knowledge of $%
\Gamma $.
\end{description}

\noindent As discussed in Ref.~\cite{NatPhys} the previous two-way protocols
must be modified into safer hybrid formulations, $Hom^{1,2}$ and $Het^{1,2}$%
, where two-way QC is randomly switched with one-way QC. In the hybrid
formulation of these protocols, the previous two-way QC is called the
\textquotedblleft ON configuration\textquotedblright\ and must be randomly
switched with an \textquotedblleft OFF configuration\textquotedblright . In
the OFF\ configuration, Alice simply detects the reference mode $R$ (via
homodyne or heterodyne) and sends a new reference mode $\tilde{R}$\ back to
Bob \cite{NatPhys}. In both the ON and OFF\ configurations, Alice and Bob
finally disclose part of the data in order to perform tomography of the
quantum channel. Thanks to this information, Alice and Bob can reconstruct
Eve's attack. In particular, they are able to understand if Eve is
exploiting quantum and/or classical correlations between the forward and
backward paths (two-mode attacks). If this is not the case (one-mode
attacks), they use the ON\ instances to generate the secret key. Otherwise,
they can use the OFF instances \cite{NatPhys}.

\section{Security against asymmetric Gaussian attacks}

Notice that in Ref.~\cite{NatPhys}, the quantitative cryptoanalysis is
restricted to one-mode Gaussian attacks, where independent and identical
Gaussian interactions affect the forward and backward channels of the
two-way quantum communication. Here, we study an extension of this analysis
by considering attacks where the Gaussian interactions are independent but
no longer identical. By independent interactions we mean interactions which
are incoherent, i.e., void of quantum correlations. However, since these
interactions are generally different, they can be classically correlated,
i.e., specified by correlated parameters \cite{Correlations}. A general
analysis of these \textquotedblleft asymmetric Gaussian
attacks\textquotedblright\ is very difficult. For this reason, we consider
only specific classes which are constructed using entangling cloners and/or
lossy channels. Further, our cryptoanalysis concerns direct reconciliation
only. Under these assumptions we are able to prove that the ON configuration
of the hybrid protocols (two-way QC) still provides a security enhancement.

Let us study the security of the hybrid protocol $Hom^{1,2}$ against
(individual) asymmetric Gaussian attacks which are based on the combination
of entangling cloners. Let us assume that Alice and Bob generate the secret
key using the ON configuration of the protocol, i.e., the two-way QC. If we
label the forward and backward channels by $i=1,2$ respectively, then we
must combine two entangling cloners with free parameters $T_{1},W_{1}$ and $%
T_{2},W_{2}$ (i.e., added noises $\Sigma _{1}$ and $\Sigma _{2}$). By
homodyning their outputs in the correct basis, Eve constructs an optimal
estimator $Q^{(E)}$ [or $P^{(E)}$] of Alice's variable. This enables her to
eavesdrop the mutual information $I_{AE}=(1/2)\ln (V/V_{A|E})$, where the
conditional variance $V_{A|E}\equiv V_{Q|Q^{(E)}}=V_{P|P^{(E)}}$ quantifies
Eve's remaining uncertainty on Alice's variable. Similarly, Bob's estimator $%
Q^{(B)}$ [or $P^{(B)}$] leaves him with a conditional variance $%
V_{A|B}\equiv V_{Q|Q^{(B)}}=V_{P|P^{(B)}}$. For $T_{i}\neq 0,1$ and $%
V\rightarrow +\infty $, one derives%
\begin{eqnarray}
V_{A|B} &=&\frac{T_{2}(1-T_{1})W_{1}+(1-T_{2})W_{2}}{T_{2}}~,
\label{Var_Attack} \\
V_{A|E} &=&\frac{T_{2}(1-T_{1})W_{2}^{-1}+(1-T_{2})W_{1}^{-1}}{%
(1-T_{1})(1-T_{2})}~.  \label{Var_Attack2}
\end{eqnarray}%
Let us consider the minimum of $V_{A|B}V_{A|E}$, so that Eve minimizes her
perturbation of the channel ($V_{A|B}$) while maximizing the acquired
information (inverse of $V_{A|E}$). Such a product takes the minimum value $%
V_{A|B}V_{A|E}=4$ for
\begin{equation}
W_{2}=1\text{ \ and \ }T_{2}=[1+(1-T_{1})W_{1}]^{-1}.  \label{T2}
\end{equation}%
\newline
The latter condition corresponds to considering an entangling cloner with
free parameters $(T_{1},W_{1})$ on the forward channel, followed by a beam
splitter with a classically \emph{correlated}\ transmission $%
T_{2}=f(T_{1},W_{1})$ on the backward channel. In order to derive the
security threshold we must impose the condition $I_{AB}=I_{AE}$ which is
equivalent to $V_{A|B}=V_{A|E}$. By using Eqs.~(\ref{Var_Attack}), (\ref%
{Var_Attack2}) and~(\ref{T2}), we get $W_{1}=(1-T_{1})^{-1}$ and $T_{2}=1/2$%
. These parameters characterize the curve of the threshold attacks which
have total noise equal to
\begin{equation}
\Sigma \equiv \Sigma _{1}+\Sigma _{2}=1+T_{1}^{-1}\text{~.}  \label{SIGMA}
\end{equation}%
It follows that the security threshold of $Hom^{1,2}$ satisfies $\Sigma >2$,
to be compared with the security threshold $\Sigma =1$ of the corresponding
one-way protocol $Hom^{1}$. In other words, when the communication channel
is too noisy for one-way protocols, it can still be used by two-way
protocols to generate a secret key.

In order to support further this \textquotedblleft
superadditivity\textquotedblright , we also study the case of asymmetric
lossy-channel attacks where the two paths of QC\ are attacked by two beam
splitters with different (correlated) transmissions $T_{1}$ and $T_{2}$.
Once the correct basis is disclosed by Bob, Eve homodynes their output ports
$E_{1}^{\prime }$ and $E_{2}^{\prime }$ to infer the signal (in the
individual version of the attack). Since two beam splitters are two
entangling cloners with $W_{1}=W_{2}=1$, from Eqs.~(\ref{Var_Attack}) and~(%
\ref{Var_Attack2}) we get
\begin{equation}
V_{A|B}=\frac{1-T_{1}T_{2}}{T_{2}}~,
\end{equation}%
and
\begin{equation}
V_{A|E}=\frac{1-T_{1}T_{2}}{(1-T_{1})(1-T_{2})}~.
\end{equation}%
Then, from $V_{A|B}=V_{A|E}$ we get the threshold curve for this kind of
attack, i.e.,
\begin{equation}
T_{2}=(1-T_{1})(1-T_{2})~.  \label{T_soglia}
\end{equation}%
The total transmission $T\equiv T_{1}T_{2}$ has a maximum equal to $3-2\sqrt{%
2}$ on this curve. Such a value corresponds to a threshold of about $7.65$dB
of losses, to be compared with the $3$dB limit of the one-way protocol.

More strongly, we prove that this threshold remains the same even when we
change the nature of the lossy-channel attack from individual to collective.
In the collective attack, Eve keeps her output probes until the end of the
protocol, when she exploits all the classical information exchanged by Alice
and Bob to perform a final coherent measurement on all her probes. In such a
case, the key rate is bounded by $I_{AB}-\chi _{E}$ where $\chi _{E}$ is the
Holevo information of the ensemble $\rho _{E}=\int G(Q)\rho _{E}(Q)dQ$ (here
$\rho _{E}(Q)$ is Eve's conditional state, while $G(Q)$ is a Gaussian with
variance $\left\langle Q^{2}\right\rangle =V$). For $T_{i}\neq 0,1$ and $%
V\rightarrow +\infty $, one can prove (see Appendix) that
\begin{equation}
\chi _{E}=\frac{1}{2}\ln \left[ \frac{V(1-T_{1})(1-T_{2})}{1-T_{1}T_{2}}%
\right] ~.  \label{Chi_Eve_lossy}
\end{equation}%
In the same limit, Alice and Bob's mutual information is given by
\begin{equation}
I_{AB}=\frac{1}{2}\ln \left( \frac{V}{V_{A|B}}\right) \rightarrow \frac{1}{2}%
\ln \left( \frac{T_{2}V}{1-T_{1}T_{2}}\right) ~.  \label{I_AB_lossy}
\end{equation}%
As a consequence, the threshold condition $I_{AB}=\chi _{E}$ gives the same
curve of Eq.~(\ref{T_soglia}), so that the security threshold remains 7.65dB.

Let us now study the security of the hybrid protocol $Het^{1,2}$ against
(collective) asymmetric lossy-channel attacks. Let us assume again that
Alice and Bob use the ON configuration to generate the secret key. For $%
T_{i}\neq 0,1$ and $V\rightarrow +\infty $, one derives $I_{AB}=\ln
(T_{2}V/2)$ while Eve's accessible information is bounded by%
\begin{equation}
\chi _{E}=\ln \left[ \frac{eV(1-T_{1})(1-T_{2})}{2(1-T_{1}T_{2})}\right] ~.
\label{Chi_E}
\end{equation}%
Then, from the condition $I_{AB}=\chi _{E}$, one finds the curve
\begin{equation}
T_{2}(1-T_{1}T_{2})=e(1-T_{1})(1-T_{2})~.
\end{equation}%
On this curve the total transmission $T\equiv T_{1}T_{2}$ has a maximum
equal to $e(e+4)^{-1}$, corresponding to about 3.93dB. Such a value must be
compared with the threshold of 1.4dB found for the corresponding $Het$
protocol \cite{LCcollective}. Notice that if we allow Bob to perform
coherent measurements (on all his states) in order to retrieve Alice's
signal $(Q,P)$, then we can reach the same security performances of the $%
Hom^{1,2}$ protocol. In such a case, in fact, Bob can asymptotically
approach the accessible information%
\begin{equation}
\chi _{B}=\ln \left[ \frac{eT_{2}V}{2(1-T_{1}T_{2})}\right] ~,
\label{Chi_Bob}
\end{equation}%
for $V\rightarrow +\infty $ and $T_{i}\neq 0,1$ (see Appendix). From the
threshold condition $\chi _{B}-\chi _{E}=0$, we then get the same curve of
Eq.~(\ref{T_soglia}) and, therefore, the same security threshold of $7.65$dB
as $Hom^{1,2}$.

\section{Conclusion}

In conclusion, multi-way quantum cryptography represents a new environment
to develop and extend quantum key distribution\ protocols. In this paper we
have studied the security of two-way protocols against Gaussian attacks
which are asymmetric between the two paths of the quantum communication. We
have shown that, even in the presence of these asymmetric strategies, the
superadditivity of the two-way quantum communication is preserved in direct
reconciliation. In particular, this is true for an important class of
asymmetric Gaussian attacks, i.e., the asymmetric lossy-channel attacks.
These analyses represent further steps to assess the security of two-way
schemes in the context of continuous variable quantum cryptography.

\section{Acknowledgements}

The research of S. Pirandola was supported by a Marie Curie Outgoing
International Fellowship within the 6th European Community Framework
Programme. S.L. was supported by the W.M. Keck center for extreme quantum
information processing (xQIT).

\appendix

\section{Appendix}

\subsection{Estimators and remote state preparation}

Consider the general scenario where Alice and Bob share two modes $A$ and $B$%
, whose quadratures $\vec{\xi}\equiv (Q_{A},P_{A},Q_{B},P_{B})$ satisfy the
canonical commutation relations $[\xi _{l},\xi _{m}]=2i\mathbf{J}_{lm}$,
where
\begin{equation}
\mathbf{J}=\left(
\begin{array}{cc}
0 & 1 \\
-1 & 0%
\end{array}%
\right) \oplus \left(
\begin{array}{cc}
0 & 1 \\
-1 & 0%
\end{array}%
\right) ~.
\end{equation}%
Suppose that modes $A$ and $B$ are described by a bipartite Gaussian state $%
\rho _{AB}$, with zero displacement $d\equiv \langle \vec{\xi}\rangle =0$
and correlation matrix (CM) $\mathbf{V}$, with generic entries $\mathbf{V}%
_{lm}\equiv \langle \xi _{l}\xi _{m}+\xi _{m}\xi _{l}\rangle /2$. The CM $%
\mathbf{V}$ is a real and symmetric matrix that must satisfy the Heisenberg
principle%
\begin{equation}
\mathbf{V}+i\mathbf{J}\geq 0~,  \label{Heis_princ_CM}
\end{equation}%
taking the form $\left\langle Q_{A}^{2}\right\rangle \left\langle
P_{A}^{2}\right\rangle \geq 1$ for the diagonal elements. All the quantum
and/or classical correlations between the modes are described by the CM
which we assume to be completely known to the parties.

Then, suppose that Alice homodynes mode $A$ and Bob homodynes mode $B$, both
of them projecting onto the same quadrature, e.g., $Q$. Thanks to the shared
correlations, Alice is able to infer Bob's outcome $Q_{B}$ from the outcome $%
Q_{A}$ of her measurement \cite{Estimators}. In fact, from $Q_{A}$, Alice
can construct the optimal estimator $Q_{B}^{(A)}\equiv \kappa Q_{A}$ of the
variable $Q_{B}$, where $\kappa \equiv \langle Q_{A}Q_{B}\rangle \langle
Q_{A}^{2}\rangle ^{-1}$ is directly computable from the CM. After her
estimation, Bob's variable $Q_{B}$, with initial variance $V_{Q_{B}}\equiv
\langle Q_{B}^{2}\rangle $, will be reduced to the conditional variable $%
Q_{B|A}\equiv Q_{B}-Q_{B}^{(A)}$ with conditional variance
\begin{eqnarray}
V_{Q_{B}|Q_{A}} &\equiv &\langle Q_{B|A}^{2}\rangle =\left\langle
Q_{B}^{2}\right\rangle -\frac{\left\langle Q_{B}^{(A)}Q_{B}\right\rangle ^{2}%
}{\left\langle Q_{B}^{(A)2}\right\rangle }  \notag \\
&=&\left\langle Q_{B}^{2}\right\rangle -\frac{\left\langle
Q_{A}Q_{B}\right\rangle ^{2}}{\left\langle Q_{A}^{2}\right\rangle }\text{~}.
\label{min_Variance}
\end{eqnarray}%
Thanks to Alice's estimation, the Shannon entropy $H(Q_{B})=(1/2)\ln
V_{Q_{B}}$ of Bob's variable has been reduced to the conditional entropy $%
H(Q_{B}|Q_{A})=(1/2)\ln V_{Q_{B}|Q_{A}}$. Therefore, the mutual information
of Alice and Bob will be given by
\begin{equation}
I(Q_{B},Q_{A})=H(Q_{B})-H(Q_{B}|Q_{A})=\frac{1}{2}\ln \frac{V_{Q_{B}}}{%
V_{Q_{B}|Q_{A}}}~.  \label{mutual_info}
\end{equation}%
Now, if we do not consider Bob's measurement, Alice's local measurement
corresponds to a remote state preparation at Bob's site. In fact, her
measurement simply corresponds to a Gaussian quantum operation that projects
Bob's mode onto a Gaussian state, centered at the point $\{Q_{B}^{(A)},0\}$
of phase-space, and with uncertainties equal to $V_{Q_{B}|Q_{A}}$ of Eq.~(%
\ref{min_Variance}) and $V_{P_{B}|P_{A}}\geq V_{Q_{B}|Q_{A}}^{-1}$. More
generally, Alice can remotely prepare a Gaussian state by making a joint
measurement of her quadratures $Q_{A}$ and $P_{A}$. For instance, she can
perform a heterodyne detection by inserting her mode $A$ into a balanced
beam splitter and, then, detecting the quadratures $Q_{+}$ and $P_{-}$ of
the output modes `$\pm $' (see Fig.~\ref{Attack1}). From the outcomes $%
(Q_{+},P_{-})$, Alice can construct two optimal estimators $Q_{B}^{(+)}=\xi
_{+}Q_{+}$\ and $P_{B}^{(-)}=\xi _{-}P_{-}$,\ so that Bob's variables $Q_{B}$%
\ and $P_{B}$\ are reduced to the conditional ones $Q_{B|+}\equiv
Q_{B}-Q_{B}^{(+)}$ and $P_{B|-}\equiv P_{B}-P_{B}^{(-)}$, with conditional
variances $V_{Q_{B}|Q_{+}}$ and $V_{P_{B}|P_{-}}$ [computable from the CMs
of $\rho _{+B}$ and $\rho _{-B}$\ according to Eq.~(\ref{min_Variance})]. In
other words, Alice remotely prepares a Gaussian state centered at $%
\{Q_{B}^{(+)},P_{B}^{(-)}\}$ with uncertainties $V_{Q_{B}|Q_{+}}$ and $%
V_{P_{B}|P_{-}}$ In particular, if the shared Gaussian state $\rho _{AB}$ is
an EPR\ source with variance $V$ [see Eq.~(\ref{CM_EPR})] then $%
V_{Q_{B}|Q_{+}}=V_{P_{B}|P_{-}}=1$ and, therefore, Alice prepares a coherent
state $\left\vert \gamma \right\rangle $\ with amplitude $\gamma
=[Q_{B}^{(+)}+iP_{B}^{(-)}]/2$. Due to the probabilistic behavior of the
measurement, the amplitude $\gamma $ represents a complex random variable
over many instances of the process. Such a variable follows a Gaussian
distribution with zero mean and second moments given by $\langle
Q_{B}^{(+)2}\rangle =\langle P_{B}^{(-)2}\rangle =V-1$ and $\langle
Q_{B}^{(+)}P_{B}^{(-)}\rangle =0$. Therefore, the\ physical scheme where
Alice and Bob share an EPR\ source with variance $V$ and Alice heterodynes
her mode is equivalent to a black-box where Alice prepares a coherent state
whose amplitude is modulated by a Gaussian distribution with variance $V-1$.
In this sense, prepare and measure schemes using coherent states are
equivalent to EPR\ schemes.

\subsection{Computation of the relevant entropies}

Consider the case of a collective and asymmetric lossy-channel attack
against the protocol $Hom^{1,2}$ in the ON configuration\textbf{.} This
means that Eve exploits two beam-splitters of (correlated) transmissions $%
T_{1},T_{2}$ and performs a final coherent measurement on all her probes.
Eve's output modes $E_{1}^{\prime }$, $E_{2}^{\prime }$ are described by a
state $\rho _{E}(Q)$ which is conditioned to Alice's encoding $Q$. On
average, Eve gets an ensemble $\rho _{E}=\int G(Q)\rho _{E}(Q)dQ$, where $%
G(Q)$ is a Gaussian distribution with variance $\left\langle
Q^{2}\right\rangle =V$. The Holevo information of Eve is then equal to $\chi
_{E}=S_{E}-S_{E|A}$, where $S_{E}$ and $S_{E|A}$ are the Von Neumann
entropies of $\rho _{E}$ and $\rho _{E}(Q)$ (computable from the CMs $%
\mathbf{V}_{E}$ and $\mathbf{V}_{E|A}$\ of the corresponding Gaussian
states). One can prove that $\mathbf{V}_{E}=\mathbf{V}_{12}\oplus \mathbf{I}%
\oplus \mathbf{I}$, where
\begin{equation}
\mathbf{V}_{12}=\left(
\begin{array}{cc}
\mu _{1}\mathbf{I}~ & ~\theta \mathbf{I} \\
\theta \mathbf{I}~ & ~~\mu _{2}\mathbf{I}+\mathbf{\Omega }\left( V,V\right)%
\end{array}%
\right) ~,  \label{CM_Eve}
\end{equation}%
with
\begin{eqnarray}
\mu _{1} &\equiv &T_{1}+(1-T_{1})V~, \\
\mu _{2} &\equiv &1+T_{1}(1-T_{2})(V-1)~, \\
\theta &\equiv &\sqrt{T_{1}(1-T_{1})(1-T_{2})}\left( V-1\right) ~,
\end{eqnarray}%
and%
\begin{equation}
\mathbf{\Omega }(V_{Q},V_{P})\equiv (1-T_{2})\left(
\begin{array}{cc}
V_{Q} & 0 \\
0 & V_{P}%
\end{array}%
\right) ~.
\end{equation}%
The Von Neumann entropy $S_{E}$ of the Gaussian state $\rho _{E}$ can be
computed from the symplectic eigenvalues \cite{Entropia} $\nu _{k}$ of the
CM $\mathbf{V}_{E}$ according to the formula%
\begin{equation}
S_{E}=\sum_{k=1}^{4}g(\nu _{k})~,
\end{equation}%
where
\begin{equation}
g(x)\equiv \left( \frac{x+1}{2}\right) \ln \left( \frac{x+1}{2}\right)
-\left( \frac{x-1}{2}\right) \ln \left( \frac{x-1}{2}\right) ~.
\end{equation}%
Note that, for $x\rightarrow +\infty $, the latter function adopts the
asymptotic expression \cite{NatPhys,LCcollective}
\begin{equation}
g(x)\rightarrow 1+\ln (x/2)+O(x^{-1})~.
\end{equation}%
Since $\mathbf{V}_{E}=\mathbf{V}_{12}\oplus \mathbf{I}\oplus \mathbf{I}$, we
have that%
\begin{equation}
\nu _{1}=\nu _{-}~,~\nu _{2}=\nu _{+}~,~\nu _{3}=\nu _{4}=1~,
\end{equation}%
where $\nu _{\pm }$ are the symplectic eigenvalues of $\mathbf{V}_{12}$. For
non trivial attacks ($T_{i}\neq 0,1$) and high modulation ($V\rightarrow
+\infty $), the symplectic eigenvalues $\nu _{\pm }$ become proportional to $%
V$. In particular, one has
\begin{equation}
\nu _{+}\nu _{-}=\sqrt{\det \mathbf{V}_{12}}\rightarrow
(1-T_{1})(1-T_{2})V^{2}~.
\end{equation}%
In the same limit, the entropy becomes%
\begin{eqnarray}
S_{E} &=&g(\nu _{-})+g(\nu _{+})\rightarrow 2+\ln \left[ \frac{1}{4}%
\lim_{V\rightarrow +\infty }\sqrt{\det \mathbf{V}_{12}}\right]  \notag \\
&=&2+\ln \left[ \frac{V^{2}}{4}(1-T_{1})(1-T_{2})\right] ~.
\end{eqnarray}

The conditional entropy $S_{E|A}$ can be computed from the symplectic
eigenvalues of the matrix $\mathbf{V}_{E|A}$. It is easy to verify that $%
\mathbf{V}_{E|A}$ can be derived from $\mathbf{V}_{E}$ by substituting $%
\mathbf{\Omega }(0,V)$ for $\mathbf{\Omega }(V,V)$ in Eq.~(\ref{CM_Eve}).
Then, repeating the previous steps, one finds
\begin{equation}
S_{E|A}\rightarrow 2+\frac{1}{2}\ln \left[ \frac{V^{3}}{16}%
(1-T_{1})(1-T_{2})(1-T_{1}T_{2})\right] ~,
\end{equation}%
so that $\chi _{E}$ is equal to Eq.~(\ref{Chi_Eve_lossy}).

Consider now a collective and asymmetric lossy-channel attack against the
protocol $Het^{1,2}$ in the ON configuration. Eve's entropy $S_{E}$ is the
same as before, while the partial entropy $S_{E|A}$\ is now conditioned to
both of Alice's variables $Q$\ and $P$. This entropy can be derived from the
conditional CM $\mathbf{V}_{E|A}$, which is computed from $\mathbf{V}_{E}$
by substituting $\mathbf{\Omega }(0,0)$ for $\mathbf{\Omega }(V,V)$ in Eq.~(%
\ref{CM_Eve}). For $T_{i}\neq 0,1$ and taking $V\rightarrow +\infty $, one
finds
\begin{equation}
S_{E|A}\rightarrow 1+\ln \left[ \frac{V}{2}(1-T_{1}T_{2})\right] ~,
\end{equation}%
so that the Holevo information $\chi _{E}$ is equal to Eq.~(\ref{Chi_E}).
Now, let us allow Bob to perform a coherent measurement on all his states,
in order to retrieve the full signal $\gamma =(Q+iP)/2$ encoded by Alice.
Bob's modes $r$ and $B^{\prime }$ are described by a state $\rho _{B}(\gamma
)$ which is conditioned to Alice's encoding $\gamma $. On average, Bob gets
an ensemble $\rho _{B}=\int G(\gamma )\rho _{B}(\gamma )d^{2}\gamma $, where
$G(\gamma )$ is a Gaussian distribution with $\left\langle
Q^{2}\right\rangle =\left\langle P^{2}\right\rangle =V$ and $\left\langle
QP\right\rangle =0$. The Bob's Holevo information is then equal to $\chi
_{B}=S_{B}-S_{B|A}$, where the two Von Neumann entropies $S_{B}$ and $%
S_{B|A} $ are computable from the CMs of $\rho _{B}$ and $\rho _{B}(\gamma )$
exactly as before. One can verify that $\rho _{B}$ has the CM%
\begin{equation}
\mathbf{V}_{B}=\left(
\begin{array}{cc}
V\mathbf{I}~ & ~\varphi \mathbf{Z} \\
\varphi \mathbf{Z}~ & ~[\varsigma +\Omega (V)]\mathbf{I}%
\end{array}%
\right) ~,  \label{V_B}
\end{equation}%
where
\begin{eqnarray}
\varphi &\equiv &\sqrt{T_{1}T_{2}(V^{2}-1)}~, \\
\varsigma &\equiv &1+T_{1}T_{2}(V-1)~,
\end{eqnarray}%
and%
\begin{equation}
\Omega (V)=T_{2}V~.
\end{equation}%
For $T_{i}\neq 0,1$ and $V\rightarrow +\infty $, the symplectic eigenvalues
of $\mathbf{V}_{B}$\ become proportional to $V$ and the entropy becomes%
\begin{equation}
S_{B}\rightarrow 2+\ln \left[ \frac{1}{4}\lim_{V\rightarrow +\infty }\sqrt{%
\det \mathbf{V}_{B}}\right] =2+\ln \left( \frac{T_{2}V^{2}}{4}\right) ~.
\label{S_Bob}
\end{equation}%
Then, the CM $\mathbf{V}_{B|A}$ of $\rho _{B}(\gamma )$ can be computed by
substituting $\Omega (0)$ for $\Omega (V)$ in Eq.~(\ref{V_B}). In the usual
limit, we have $\nu _{-}=1$ and $\nu _{+}\rightarrow V(1-T_{1}T_{2})$, so
that
\begin{equation}
S_{B|A}=g(\nu _{+})\rightarrow 1+\ln \left[ \frac{V}{2}(1-T_{1}T_{2})\right]
~.  \label{S_Bob_A}
\end{equation}%
From Eqs.~(\ref{S_Bob}) and~(\ref{S_Bob_A}), one easily gets Eq.~(\ref%
{Chi_Bob}) for Bob's Holevo information.


\begin{thebibliography}{99}
\bibitem{CVbook} S. L. Braunstein, and A. K. Pati, \textit{Quantum
Information Theory with Continuous Variables}, Kluwer Academic, Dordrecht,
2003; S. L. Braunstein, and P. van Loock, \textquotedblleft Quantum
information with continuous variables,\textquotedblright\ Rev. Mod. Phys.
\textbf{77}, 513 (2005).

\bibitem{Ralph1} T. C. Ralph, \textquotedblleft Continuous variable quantum
cryptography,\textquotedblright\ Phys. Rev. A \textbf{61}, 010303(R) (2000);
T. C. Ralph, \textquotedblleft Security of continuous-variable quantum
cryptography,\textquotedblright\ Phys. Rev. A \textbf{62}, 062306 (2000);\
M. D. Reid, \textquotedblleft Quantum cryptography with a predetermined key
using continuous-variable Einstein-Podolsky-Rosen
correlations,\textquotedblright\ Phys. Rev. A \textbf{62}, 062308 (2000).

\bibitem{Preskill} D. Gottesman, and J. Preskill, \textquotedblleft Secure
quantum key distribution using squeezed states,\textquotedblright\ Phys.
Rev. A \textbf{63}, 022309 (2001); S. Iblisdir, G. Van Assche, and N. J.
Cerf, \textquotedblleft Security of quantum key distribution with coherent
states and homodyne detection,\textquotedblright\ Phys. Rev. Lett\textit{.}
\textbf{93}, 170502 (2004).

\bibitem{Homo} F. Grosshans, and Ph. Grangier, \textquotedblleft Continuous
variable quantum cryptography using coherent states,\textquotedblright\
Phys. Rev. Lett. \textbf{88}, 057902 (2002).

\bibitem{Hetero} C. Weedbrook \textit{et al., }\textquotedblleft Quantum
cryptography without switching,\textquotedblright\ Phys. Rev. Lett. \textbf{%
93}, 170504 (2004).

\bibitem{Homo2} F. Grosshans \textit{et al.}, \textquotedblleft Quantum key
distribution using Gaussian-modulated coherent states,\textquotedblright\
Nature \textbf{421}, 238 (2003).

\bibitem{Hetero2} A. M. Lance \textit{et al.},\textit{\ }\textquotedblleft
No-switching quantum key distribution using broadband modulated coherent
light,\textquotedblright\ Phys. Rev. Lett. \textbf{95}, 180503 (2005).

\bibitem{GaussRec} G. Van Assche \textit{et al.}, \textquotedblleft
Reconciliation of a quantum-distributed Gaussian key,\textquotedblright\
IEEE Trans. Inform. Theory \textbf{50}, 394 (2004).

\bibitem{Estimators} F. Grosshans \textit{et al.}, \textquotedblleft Virtual
entanglement and reconciliation protocols for quantum cryptography with
continuous variables,\textquotedblright\ Quant. Info. and Computation
\textbf{3}, 535 (2003).

\bibitem{QObook} D. F. Walls, and G. J. Milburn, \textit{Quantum Optics},
Springer, 1994.

\bibitem{NatPhys} S. Pirandola, S. Mancini, S. Lloyd, and S. L. Braunstein,
\textquotedblleft Quantum cryptography using two-way quantum
communication,\textquotedblright\ Nature Physics advance online publication,
11 July 2008 (arXiv:quant-ph/0611167v2).

\bibitem{Correlations} Notice that, in general, the second interaction can
be conditioned to both the parameters of the first interaction and the
outcomes of a (possible) measurement which detects the corresponding output
modes of the eavesdropper. However, in the limit of large modulation, it is
reasonable to consider \textit{universal} interactions which, therefore, are
not conditioned to any outcome. As a consequence, the classical correlations
can be reduced to correlations between the parameters of the two
interactions.

\bibitem{LCcollective} F. Grosshans, \textquotedblleft Collective attacks
and unconditional security in continuous variable quantum key
distribution,\textquotedblright\ Phys. Rev. Lett. \textbf{94}, 020504
(2005); M. Navascu\'{e}s, and A. Ac\'{\i}n, \textquotedblleft Security
bounds for continuous variables quantum key distribution,\textquotedblright\
Phys. Rev. Lett. \textbf{94}, 020505 (2005).

\bibitem{Entropia} A. S. Holevo \textit{et al.}, \textquotedblleft Capacity
of quantum Gaussian channels,\textquotedblright\ Phys. Rev. A \textbf{59},
1820 (1999).
\end{thebibliography}
\end{document}